\documentclass[twocolumn,
prd,secnumarabic,amsmath,amssymb,superscriptaddress,
nofootinbib]{revtex4-2}

\usepackage{graphicx} 
\usepackage{xcolor}
\newcommand{\mL}{{\mathcal{L}}}
\newcommand{\mE}{{\mathcal{E}}}
\newcommand{\mB}{{\mathcal{B}}}
\newcommand{\mD}{{\mathcal{D}}}
\newcommand{\mH}{{\mathcal{H}}}
\newcommand{\mP}{{\mathcal{P}}}
\newcommand{\mM}{{\mathcal{M}}}
\newcommand{\req}[1]{Eq.\,(\ref{#1})} 
\newcommand{\rs}[1]{section~\ref{#1}} 
\newcommand{\rf}[1]{Fig.\,\ref{#1}} 

\begin{document}

\title{{\normalsize\sc  Dedicated to Professor Iwo Białynicki-Birula on His 90th Birthday}
\\ \vspace{0.75cm}
Improving Euler-Heisenberg-Schwinger effective action with dressed photons}

\author{Stefan Evans}
\email{evanss@arizona.edu}
\affiliation{Department of Physics, The University of Arizona, Tucson, AZ 85721, USA}
\affiliation{Helmholtz-Zentrum Dresden-Rossendorf, Bautzner Landstraße 400, 01328 Dresden, Germany}
\author{Johann Rafelski}
\affiliation{Department of Physics, The University of Arizona, Tucson, AZ 85721, USA}

\begin{abstract}
We implement a longstanding proposal by Weisskopf to apply virtual polarization corrections to the in/out external fields in study of the Euler-Heisenberg-Schwinger effective action. Our approach requires distinguishing the electromagnetic and polarization fields based on mathematical tools developed by Białynicki-Birula, originally for the Born-Infeld action. Our solution is expressed as a differential equation where the one-loop effective action serves as input. As a first result of our approach, we recover the higher-order one-cut reducible loop diagrams discovered by Gies and Karbstein. 
\end{abstract}

\maketitle

\section{Introduction}

Victor Weisskopf in 1936~\cite{Weisskopf:1996bu, Weisskopf:1996bu2} suggested and attempted an improvement to the derivation of the Heisenberg-Euler effective action~\cite{Heisenberg:1935qt}; for further insights see later work by Schwinger~\cite{Schwinger:1951nm} and the review by Dunne~\cite{Dunne:2004nc}. Weisskopf considered that the polarization of the vacuum should be `fortw\"ahrend' (everlasting) and thus photons should contain the polarization effects already present in a self-consistent manner. In present-day language the class of diagrams he envisaged requires summation of one-cut reducible loop diagrams, {\it i.e.}, photons dressed by one-loop Euler-Heisenberg-Schwinger (EHS) action. In this work we present a path to a solution of this problem and give examples using constant homogeneous electromagnetic (EM) fields.

At first, the reducible loop diagram contributions to quantum electrodynamic (QED) effective action were assumed to vanish in the (infrared) {\it i.e.\/} the constant field limit: Ritus~\cite{Ritus:1975cf} claimed that as photon momentum $k\to0$, the pertinent  two-loop diagrams vanish in view of the current $\propto k^2$. However, Gies and Karbstein~\cite{Gies:2016yaa} discovered that the pole of the virtual photon propagator ($\propto 1/k^2$) perfectly cancels the vanishing current in the quasi-constant EM field limit. This study of the nonvanishing two-loop reducible diagram corrections to EHS effective action was extended via further perturbative summation to higher-order loops~\cite{Karbstein:2017gsb, Karbstein:2019wmj, Karbstein:2021gdi}, to scalar~\cite{Edwards:2017bte} and spinor propagators~\cite{Ahmadiniaz:2017rrk}, and to a more general class of field configurations~\cite{Ahmadiniaz:2019nhk}.

In this work we demonstrate the connection between the Weisskopf conjecture and these reducible loop diagrams discovered in the present day field-theoretical context. We implement a classical polarization approach for summing virtual photon excitations in the infrared limit. By dressing the external field with polarization corrections at the start of the derivation of EHS action, we recover the two-loop result of Gies and Karbstein~\cite{Gies:2016yaa}. 

A key input into our nonperturbative solution is a class of Legendre transforms of nonlinear EM actions formulated by Białynicki-Birula~\cite{BialynickiBirula:1984tx},  allowing to transform the nonlinear EHS action --- a function of EM fields $\mL_{\mathrm{1}}(\mE,\mB)$ --- into an expression employing superposable  fields $\mD,\mH$. In this step  we can insert polarization corrections to dress the external fields. Lastly, we inverse the Legendre transform to return to an effective action formulation in terms of EM fields.

In~\rs{WeisskopfReview}, we develop an approach for implementing Weisskopf's proposal to improve the EHS result, based on polarization corrections to the external fields. We implement the corrections in~\rs{3steps}, using the Legendre transformed EHS action, and apply our theoretical result to the case of pure electric fields. In~\rs{GiesResult} we recover the two-loop effective action of Gies and Karbstein. Extension to higher-order loop contributions is straightforward, as we show with the three-loop action as an example. We believe that our approach can be applied to extend any one-loop effective action in the same everlasting manner, including the case of special interest, the strongly-interacting vacuum structure.

\section{Implementing Weisskopf}
\label{WeisskopfReview}

\subsection{Nonlinear EM action overview}

We consider a general expression for EM effective action in the infrared external field limit (photon momentum $k\to0$):
\begin{align}
\label{LEMfirst}
 \mL_{\mathrm{M+1}}(\mE,\mB)=&\;
\frac{\mE^2-\mB^2}2+\mL_{\mathrm{1}}(\mE,\mB)
\;,
\end{align}
where subscript M+1 denotes the Maxwell plus one-loop EHS contributions to the action. The EM fields $\mE,\mB$ are generated by the 4-potential $A^\mu$ governing the Lorentz force as $F^{\mu\nu}=\partial^\mu A^\nu-\partial^\nu A^\mu$, and are related to the superposable fields $\mD,\mH$ governing Maxwell equations with sources as:
\begin{align}
\label{DH}
\mD(\mE,\mB)=&\;
\frac{\partial \mL_{\mathrm{M+1}}}{\partial\mE}=\mE+ \frac{\partial \mL_{\mathrm{1}}}{\partial\mE}
\;,
\\ \nonumber
\mH(\mE,\mB)=&\;
-\frac{\partial \mL_{\mathrm{M+1}}}{\partial\mB}=\mB- \frac{\partial \mL_{\mathrm{1}}}{\partial\mB}
\;.
\end{align}
The nonlinear response of the vacuum thus distinguishes $\mE,\mB$ from these superposable fields:
\begin{align}
\label{EB}
\mE\equiv&\;\mD(\mE,\mB)-\mP(\mE,\mB)\;,
\\ \nonumber
\mB\equiv&\;\mH(\mE,\mB)+\mM(\mE,\mB)
\;,
\end{align}
where the polarization fields $\mP,\mM$ render the EM fields $\mE,\mB$ non-superposable. This distinction will be necessary in order to implement Weisskopf's proposal to dress the externally applied EM fields.

All the relevant expressions for effective action in terms of EM and superposable fields are shown in table~\ref{allLandU}. The auxiliary quantity $U$ is obtained from $\mL$ by Legendre transform, as we will describe below.
\begin{table}[h] 
\begin{tabular}{|c|c|c|} 
\hline
& Lagrange form & Auxiliary form   \\
\hline\hline
EHS & $\mL_{\mathrm{1}}(\mE,\mB)$ & $U_{\mathrm{1}}(\mD,\mH)$ \\ 
\hline
Maxwell+EHS & $\mL_{\mathrm{M+1}}(\mE,\mB)$ & $U_{\mathrm{M+1}}(\mD,\mH)$  \\
\hline
Dressed photons & $\mL_{\mathrm{W}}(\mE,\mB)$ & $U_{\mathrm{W}}(\mD,\mH)$  \\
\hline
Maxwell+Dressed photons & $\mL_{\mathrm{M+W}}(\mE,\mB)$ & $U_{\mathrm{M+W}}(\mD,\mH)$  \\
\hline
\end{tabular}
\caption{EHS action (first two rows), and the higher-order one-cut reducible loop action (last two rows). M+W refers to Maxwell+Weisskopf action, with Maxwell being the $(\mE^2-\mB^2)/2$ and $(\mD^2-\mH^2)/2$ contributions.
\label{allLandU}
}
\end{table}

\subsection{Reconciling EM fields with the everlasting vacuum}

In~\rf{MaterialTarget}, we show how Weisskopf's extension of EHS action works in context of in/out states: in panel (a), a photon scatters off a finite sized polarizable material medium. The asymptotic in/out states {\it i.e.} the EM fields before and after the interaction (black), are equivalent to the superposable fields ($\mE=\mD,\mB=\mH$). The screening by the medium (red) occurs inside the material target, with nonzero polarization fields $\mP,\mM$.

%
\begin{figure}[h]
\includegraphics[width=0.85\columnwidth]{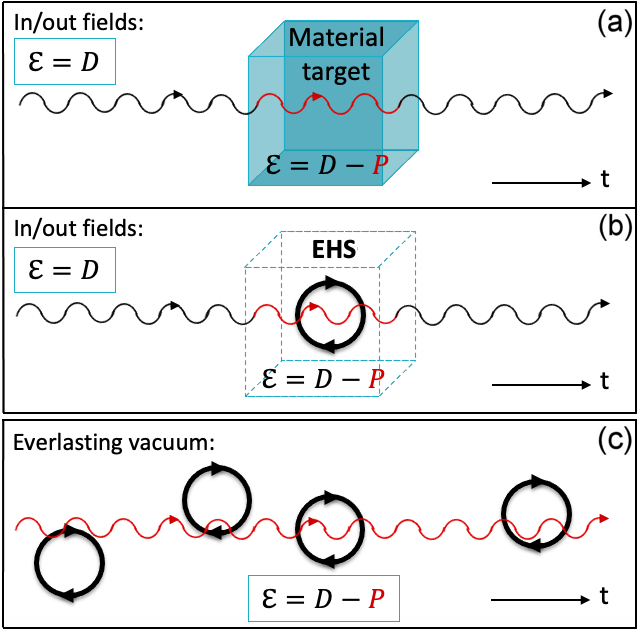}
\caption{\label{MaterialTarget}EM fields interacting with: (a) a finite sized material medium, (b) prior treatment of the perturbative EHS vacuum 
in the image of a scattering problem, (c) nonperturbative vacuum existing at all times.} 
\end{figure}
%
%

Following Weisskopf's insight that the external fields in EHS effective action see only one electron loop, we illustrate a perturbative EHS analog to the material target scattering (\rf{MaterialTarget}a). The EHS analog (\rf{MaterialTarget}b) comprises, in place of a material target, the quantum vacuum structure spanning a bounded spacetime domain sufficiently small that each photon in the external field sees only a single electron loop. Outside of this bounded region no virtual electron excitations are considered, thus the asymptotic in/out external fields are approximated as $\mE=\mD,\mB=\mH$ {\it i.e.} without polarization effects.

This perturbative approach is amended in~\rf{MaterialTarget}c. Since the vacuum structure exists at all times rather than in a bounded spacetime domain, we cannot distinguish the asymptotic in/out fields from the fields interacting with the virtual electron pairs. The polarization effects contained in fields $\mP,\mM$ are always present, and thus $\mE=\mD-\mP$ and $\mB=\mH+\mM$ throughout~\rf{MaterialTarget}c. These are the dressed fields to be implemented in the EHS action.

\section{Derivation of effective action loop summation via everlasting vacuum properties}
\label{3steps}

\subsection{Legendre transform} 

We now show how to implement polarization field $\mP,\mM$ corrections into the externally applied fields of EHS action. This cannot be done for the EHS action $\mL_{\mathrm{1}}(\mE,\mB)$ directly due to the EM fields (see~\req{EB}) being non-superposable. Thus the first step is to transform the $\mL_{\mathrm{M+1}}(\mE,\mB)$ into an auxiliary form written in terms of superposable fields $ U_{\mathrm{1}}(\mD,\mH)$, based on the Legendre transforms seen in table~\ref{LegendreT}. 
\begin{table}[h] 
\begin{tabular}{|c|c|c|c|} 
\hline
 Legendre transform & Electric field & Magnetic field  \\
\hline\hline
$\mL(\mE,\mB)=
\mE\cdot\mD-\mB\cdot\mH- U$ & 
$\mD=\partial\mL/\partial\mE$ &  $\mH=-\partial\mL/\partial\mB$  \\ 
\hline
$ U(\mD,\mH)=
\mE\cdot\mD-\mB\cdot\mH-\mL$ &
$\mE=\partial U/\partial\mD$ &  $\mB=-\partial U/\partial\mH$  \\ 
\hline
\end{tabular}
\caption{Legendre transforms and derivative expressions relating electromagnetic and superposable fields, after Ref.~\cite{BialynickiBirula:1984tx}. 
\label{LegendreT}
}
\end{table}

Carrying out the Legendre transform of the EM action~\req{LEMfirst}:
\begin{align}
\label{step1a}
 U_{\mathrm{M+1}}(\mD,\mH)
=&\;
\mE(\mD,\mH)\cdot\mD-\mB(\mD,\mH)\cdot\mH
\\ \nonumber
&\;
-\mL_{\mathrm{M+1}}(\mE(\mD,\mH),\mB(\mD,\mH))
\;,
\end{align}
where the EM fields 
\begin{align}
\label{uxuysign}
\mE(\mD,\mH)=&\;\frac{\partial U_{\mathrm{M+1}}}{\partial\mD}\;,
\\ \nonumber
\mB(\mD,\mH)=&\;-\frac{\partial U_{\mathrm{M+1}}}{\partial\mH}
\;.
\end{align} 
Separating the nonlinear contribution we define
\begin{align}
\label{EHSdef}
 U_{\mathrm{M+1}}(\mD,\mH)\equiv
\frac{\mD^2-\mH^2}2+ U_{\mathrm{1}}(\mD,\mH)
\;,
\end{align}
distinguishing the contribution to the action, in terms of $\mD,\mH$, arising from the virtual electron interaction. Note that $U_{\mathrm{1}}(\mD,\mH)$ and $\mL_{\mathrm{1}}(\mE,\mB)$ are not the same expressions, since the superposable fields take on a different functional dependence than non-superposable EM fields.  Determining $ U_{\mathrm{1}}$ requires solving an implicit differential equation as defined in~\req{step1a} and~\req{uxuysign}. An analytic solution is available for the special case of Born-Infeld action~\cite{BialynickiBirula:1984tx,Born:1934gh,Price:2023}.

\subsection{Polarization corrections}

Only in this auxiliary form of EHS effective action, using superposable $\mD,\mH$ fields, can the asymptotic in/out fields be corrected to account for everlasting polarization fields. Where $\mD,\mH$ appear in the nonlinear part of EM action $U_{\mathrm{1}}(\mD,\mH)$ in~\req{EHSdef}, we take 
\begin{align}
\mD\to\mD-\mP(\mD,\mH)=\mE(\mD,\mH)
\;,
\end{align}
and similarly for the magnetic field 
\begin{align}
\mH\to\mH+\mM(\mD,\mH)=\mB(\mD,\mH)
\;,
\end{align}
thereby dressing the asymptotically defined EM field that any single electron loop is exposed to. The polarization fields $\mP,\mM$ introduce the one-cut reducible loop sum $U_{\mathrm{W}}(\mD,\mH)$, defined as
\begin{align}
\label{resumfirst}
 U_{\mathrm{W}}(\mD,\mH)\equiv
 U_{\mathrm{1}}(\mD-\mP,\mH+\mM)
\;.
\end{align}

Including the Maxwell term and plugging in~\req{EHSdef} and~\req{EB}, we obtain
\begin{align}
\label{resumsecond3}
 U_{\mathrm{M+W}}(\mD,\mH)\equiv&\;
\frac{\mD^2-\mH^2}2 -\frac{\mE^2(\mD,\mH)-\mB^2(\mD,\mH)}2
\nonumber \\ 
&\;
+ U_{\mathrm{M+1}}(\mE(\mD,\mH),\mB(\mD,\mH))
\;,
\end{align}
where $U_{\mathrm{M+1}}(\mE,\mB)$ follows from~\req{step1a}, with the replacements $\mD\to \mE(\mD,\mH)$ and $\mB\to \mB(\mD,\mH)$.

\subsection{Inverse Legendre transform}

As a final step, we inverse Legendre transform~\req{resumsecond3} to return to the effective action formulation as a function of EM fields $\mE,\mB$. Using the transform from table~\ref{LegendreT},
\begin{align}
\label{finalLeff}
\mL_{\mathrm{M+W}}(\mE,\mB)
\equiv&\;
\mE\cdot\mD(\mE,\mB)-\mB\cdot\mH(\mE,\mB)
\\ \nonumber
&\;
- U_{\mathrm{M+W}}(\mD(\mE,\mB),\mH(\mE,\mB))
\\ \nonumber 
=&\;
\mE\cdot\mD(\mE,\mB)-\mB\cdot\mH(\mE,\mB)+\frac{\mE^2-\mB^2}2
\\ \nonumber
&\;
-\frac{\mD^2(\mE,\mB)-\mH^2(\mE,\mB)}2
-U_{\mathrm{M+1}}(\mE,\mB)
,
\;
\end{align}
where now the derivative identities 
\begin{align}
\label{DW}
\mD(\mE,\mB)=&\;
\frac{\partial\mL_{\mathrm{M+W}}(\mE,\mB)}{\partial\mE}\;,
\\ \nonumber
\mH(\mE,\mB)=&\;-\frac{\partial\mL_{\mathrm{M+W}}(\mE,\mB)}{\partial\mB}
\;.
\end{align}

Separating the Maxwell contribution from the nonlinear vacuum contribution we define 
\begin{align}
\label{Lsumfinal}
\mL_{\mathrm{W}}(\mE,\mB)\equiv&\;
\mL_{\mathrm{M+W}}(\mE,\mB)-\frac{\mE^2-\mB^2}2
\;.
\end{align}
Combining \req{finalLeff}-\req{Lsumfinal}, we now have at our disposal a differential equation requiring input EHS, which, when solved, creates the effective action for the  summed reducible loop diagrams.

\subsection{Summary and generalized form}

To summarize, we build upon the one-loop effective action $\mL_{\mathrm{M+1}}(\mE,\mB)$ in~\req{LEMfirst} by applying:
\begin{itemize}
\item{\it Legendre transform:}
\begin{align}
\!\!\!\!\!\!\!\!\!\!\!\!\!\!\!\!\!\!\!
U_{\mathrm{M+1}}(\mD,\mH)=\frac{\mD^2-\mH^2}2+U_{\mathrm{1}}(\mD,\mH)
\;,
\end{align}
\item{\it Polarization corrections:}
\begin{align}
\;\;\;\;\;
U_{\mathrm{M+W}}(\mD,\mH)=\frac{\mD^2-\mH^2}2+U_{\mathrm{1}}(\mD-\mP,\mH+\mM)
,
\end{align}
\item{\it Inverse Legendre transform:}
\begin{align}
\!\!\!\!\!\!\!\!\!\!\!\!\!\!\!\!\!\!\!\!
\mL_{\mathrm{M+W}}(\mE,\mB)=\frac{\mE^2-\mB^2}2+\mL_{\mathrm{W}}(\mE,\mB)
\;.
\end{align}
\end{itemize}

\section{Perturbative Series for $\alpha=1/137$}
\label{GiesResult}

As an illustrative example we consider the pure electric field case to study the two-loop action of Gies and Karbstein~\cite{Gies:2016yaa}. Taking $\mB\to0$,~\req{finalLeff} becomes
\begin{align}
\label{finalLeffE}
\mL_{\mathrm{M+W}}(\mE)
=&\;
\mE\cdot\mD(\mE)+\frac{\mE^2}2
-\frac{\mD^2(\mE)}2
-U_{\mathrm{M+1}}(\mE)
\;.
\end{align}
We evaluate~\req{finalLeffE} by applying a perturbative loop expansion. 

We first write the EHS Lagrangian dependence in~\req{finalLeffE} explicitly using the Legendre transform~\req{step1a}:
\begin{align} 
\label{finalLeffEexplicit}
\mL_{\mathrm{M+W}}(\mE)
=&\;
\mE\cdot\mD(\mE)-\frac{\mD^2}2
-\frac{\partial U_{\mathrm{M+1}}(\mE)}{\partial\mE}\cdot\mE
\\ \nonumber
&\;
\!\!\!\!\!\!\!\!\!\!\!\!\!\!\!\!\!\!\!\!\!\!\!\!\!
+\frac12\Big(\frac{\partial U_{\mathrm{M+1}}(\mE)}{\partial\mE}\Big)^2
+\mL_{\mathrm{1}}\Big(\frac{\partial U_{\mathrm{M+1}}(\mE)}{\partial\mE}\Big)
+\frac{\mE^2}2
\;.
\end{align}
We take the case of small polarization corrections to the externally applied EM field:
\begin{align} 
\label{weakPol}
\frac{|\mD-\mE|}{|\mE|}\ll 1
\;.
\end{align}
Under the condition~\req{weakPol}, the leading one-loop EHS contribution dominates the higher-order loop effects. The perturbative summation of reducible diagrams to $\ell$-loop order can be written as
\begin{align} 
\label{finalLeffEpert}
\lim_{\frac{|\mD-\mE|}{|\mE|}\ll 1}
\mL_{\mathrm{M+W}}(\mE)
\equiv&\;
\mL_{\mathrm{M+1}}(\mE)
+\sum_{\ell=2}^\infty\mL_\ell(\mE)
\;,
\end{align}
where the one-loop EHS contribution is included in $\mL_{\mathrm{M+1}}(\mE)$, followed by summation over the two-loop and higher-orders.

To determine the form of loop corrections $\mL_\ell(\mE)$ in~\req{finalLeffEpert} we take the small polarization limit of the auxiliary function $U_{\mathrm{M+1}}$ defined in~\req{step1a} and differentiate with respect to $\mE$ to obtain
\begin{align} 
\label{Uapprox}
\lim_{\frac{|\mD-\mE|}{|\mE|}\ll 1}\frac{\partial U_{\mathrm{M+1}}(\mE)}{\partial\mE}
=\mE-\frac{\partial \mL_{\mathrm{1}}(\mE)}{\partial\mE}
\;.
\end{align}
Similarly for the superposable field $\mD$, 
\begin{align} 
\label{Dapprox}
\lim_{\frac{|\mD-\mE|}{|\mE|}\ll 1}\mD(\mE)
=\mE+\frac{\partial \mL_{\mathrm{1}}(\mE)}{\partial\mE}
\;.
\end{align}
Plugging~\req{Uapprox} and~\req{Dapprox} into~\req{finalLeffEexplicit},
\begin{align} 
\label{LeffEpert1}
\lim_{\frac{|\mD-\mE|}{|\mE|}\ll 1}
\mL_{\mathrm{M+W}}(\mE)
=&\;
\frac{\mE^2}2+\mL_{\mathrm{1}}\Big(\mE-\frac{\partial \mL_{\mathrm{1}}(\mE)}{\partial\mE}\Big)
\;.
\end{align}
Note that \req{LeffEpert1} shows the iterative structure of the effective action describing the higher-order loop summation. Expanding in powers of $\mL_{\mathrm{1}}$:
\begin{align} 
\label{LeffEpert2}
\mL_2(\mE)=
-\Big(\frac{\partial \mL_{\mathrm{1}}(\mE)}{\partial\mE}\Big)^2
\;,
\end{align}
the two-loop of Gies and Karbstein (see~Eq.\,(32) of~\cite{Gies:2016yaa}). The original result  in~\cite{Gies:2016yaa} contains both $\mE$ and $\mB$ contributions, expressed using derivatives with respect to the EM field tensor: $\mL_2=\frac12(\partial \mL_{\mathrm{1}}/\partial F^{\mu\nu})^2=(\partial \mL_{\mathrm{1}}/\partial \mB)^2-(\partial \mL_{\mathrm{1}}/\partial \mE)^2$, which reduces to~\req{LeffEpert2} in the pure $\mE$ limit.

To obtain the three-loop contribution, we iterate the two-loop~\req{LeffEpert2} into~\req{LeffEpert1} as a complement to $\mL_{\mathrm{1}}$ appearing in the polarization correction $\partial \mL_{\mathrm{1}}(\mE)/\partial\mE$. Expanding again in powers of $\mL_{\mathrm{1}}$, this time to third order,
\begin{align} 
\label{LeffEpert3}
\mL_3(\mE)
=\frac52\frac{\partial ^2\mL_{\mathrm{1}}(\mE)}{\partial\mE^2}
\Big(\frac{\partial \mL_{\mathrm{1}}(\mE)}{\partial\mE}\Big)^2
\;.
\end{align}
This perturbative higher-order loop summation procedure can be carried out ad infinitum  as in~\cite{Karbstein:2019wmj}, with the replacement $\mB\to-i\mE$ to recast Karbstein's original summation for $\mB$ fields in terms of $\mE$ fields.

\section{Conclusions}

We have implemented Weisskopf's proposal~\cite{Weisskopf:1996bu, Weisskopf:1996bu2} to dress the external EM fields in EHS effective action with polarization corrections. This shows that the one-cut reducible QED loop diagram summation of Gies and Karbstein~\cite{Gies:2016yaa, Karbstein:2017gsb, Karbstein:2019wmj} was indeed foretold in the work of Weisskopf. We developed a  generalized approach to summing such diagrams which can be applied to any nonlinear EM theory, with a one-loop effective action as input and in principle carried to higher-order in coupling constant as we have demonstrated evaluating the next to next order correction. 

It is important to note that we include only the one-cut reducible loop diagram contributions to effective action. A full summation includes: higher-order cut reducible diagrams and internal photon line (irreducible) loops producing {\it e.g.} anomalous magnetic moment and field-dependent mass. Irreducible contributions to the action in constant fields are well-known to two-loop order~\cite{Ritus:1975cf}, and a subset of such diagrams comprising vertex corrections enclosing a single external line --- to all orders~\cite{Diet78}. 

Rather than the conventional in/out method for computing effective action which treats the structured vacuum as bounded in spacetime akin to a finite-sized material target, our approach takes into account an everlasting vacuum structure spanning all spacetime. Finally, we remark that this work complements in the \lq opposite\rq\ direction the insight by Białynicki-Birula, Rudnicki and Wienczek~\cite{BialynickiBirula:2011eg} that the finite time duration of external fields regularizes the essential singularity seen in (the imaginary part of) the one loop EHS result in the limit of weak electrical fields.

The analytical properties of the one-loop action resurface in the higher loops as a striking interplay between real and imaginary (containing the essential singularity) parts of effective action, and between reducible and irreducible diagram contributions. Strong field asymptotics need further exploration as they are highly nontrivial, depending on which EM field invariant dominates the external EM fields ($(\mE^2-\mB^2)/2$ versus the pseudoscalar $\mE\cdot\mB$). 

To conclude: we have improved the formulation of effective action in the presence of everlasting vacuum structure. Our result connects Weisskopf's conjectured extension of EHS effective action to Gies and Karbstein's discovered higher-order reducible loop diagrams.

\section*{Acknowledgments}
This work is dedicated to Prof. Iwo Białynicki-Birula on the occasion of his 90th birthday.


\end{document}